\documentclass[10pt,a4paper]{article}
\usepackage{amsmath}
\usepackage{amssymb}
\usepackage{amsfonts}
\usepackage{amstext}
\usepackage{amsbsy}
\usepackage[mathscr]{eucal}
\usepackage{graphicx}
\usepackage{color}
\usepackage[all]{xy}
\newcommand{\beq}{\begin{equation}}
\newcommand{\eeq}{\end{equation}}
\newcommand{\beqa}{\begin{eqnarray}}
\newcommand{\eeqa}{\end{eqnarray}}
\newcommand{\ket}[1]{| #1 \rangle}
\newcommand{\bra}[1]{\langle #1 |}

\newtheorem{thm}{Theorem}[subsection]

 \newtheorem{prop}[thm]{Proposition}

 \numberwithin{equation}{subsection}

\makeindex
\title{\Large\textbf{ Different classes of quantum gates entanglers}}
\author{\textit{ Hoshang Heydari}\\
        \small\textit{Physics Department, Stockholm university 10691 Stockholm Sweden}\\
\\\small\textit{Email: hoshang@physto.se}}
\date{}
\begin{document}

\maketitle \thispagestyle{empty}

\begin{abstract}
We construct quantum gates entanglers for different classes of
multipartite states. In particular we construct entangler operators
for W and GHZ classes of multipartite states based on the
construction of  the concurrence classes. We also in detail discuss
these two classes of the quantum gates entanglers for three-partite
states.
\end{abstract}

\section{Introduction}
Entangled states are very important building blocks of
fault-tolerant quantum computer. In general,  a quantum computer is
build using quantum circuit containing wires and elementary quantum
gates. For an entangled based implementation of the quantum computer
one also needs  to construct quantum gates entanglers. It is also
important that these quantum gates entanglers are able to produce
any desire class of multipartite entangled quantum  states.
  Recently, L. Kauffman and S. Lomonaco have constructed topological quantum gate entangler
   for two-qubit state \cite{kauf}. These  topological operators are called
braiding operators that can entangle quantum states. These operators
are also unitary solution of quantum Yang-Baxter equation. We have
also recently construct quantum gate entangler for multi-qubit
states based on a topological and geometrical method \cite{Hosh7}.
In particular, we have constructed unitary operators that can
entangle multi-qubit quantum  states if they satisfy the completely
separability condition defined by the ideal of the Segre embedding.
In this paper, we will construct quantum gates entanglers for
different classes of  multipartite states based on the definition of
W and GHZ concurrence classes which is also similar to the
construction of these unitary operators \cite{Hosh5,Hosh6}.  In
section \ref{ent} we will give an introduction to the construction
of Artin braid group and Yang-Baxter equation. In section \ref{con}
we review the basic construction of concurrence classes based on
orthogonal complement of positive operator valued measure (POVM) on
quantum phase. Next, in section \ref{gen} which is also the main
part of this paper, we will construct quantum gates entanglers for
different classes of entangled multipartite states. These operators
can entangle a specific class of entangled multipartite state if
they satisfy separability conditions for W and GHZ class of states.
Finally, in section \ref{exa} we visualize these classes of quantum
gates entanglers for three-partite states. For example, we will
explicitly define the W class and GHZ class quantum gates entanglers
for such a quantum state.

\section{Quantum gate entangler based on  Artin braiding operator}\label{ent}
In this section we  will study relation between
topological and quantum entanglement by investigating the unitary
representation of Artin braid group. Here are some general
references on quantum group and low-dimensional topology
\cite{kassel,chari}. The Artin braid group $\mathrm{B}_{n}$ on $n$
strands is generated by $\{b_{n}:1\leq i\leq n-1\}$. The generators of  the group $\mathrm{B}_{n}$ satisfy the following relations
 \begin{description}
   \item[i)] $
    b_{i}b_{j}=b_{j}b_{i}$ for $|i-j|\geq 2 $
and
   \item[ii)]  $b_{i}b_{i+1}b_{i}=b_{i+1}b_{i}b_{i+1}$ for $ 1\leq i<n$.
 \end{description}
 Each generator or product of generators admits an interesting $n$-strand graphical presentations. We can also find a matrix representation of the group $\mathrm{B}_{n}$  in the resulting space $\mathcal{V}^{\otimes
m}=\mathcal{V}\otimes\mathcal{V}\otimes\cdots\otimes\mathcal{V}$,
where $\mathcal{V}$ is a complex vector space. Now, for two strands
braid there is associated an operator
$R:\mathcal{V}\otimes\mathcal{V}\longrightarrow\mathcal{V}\otimes\mathcal{V}$.
Moreover, let $\mathcal{I}$ be the identity operator on
$\mathcal{V}$. Then, the quantum Yang-Baxter equation is defined by
\begin{equation}\label{YB}
    (R\otimes \mathcal{I})(\mathcal{I}\otimes R)(R\otimes
    \mathcal{I})=(\mathcal{I}\otimes R)(R\otimes \mathcal{I})(\mathcal{I}\otimes
    R).
\end{equation}
The Yang-Baxter equation represents the fundamental topological
relation in the Artin braid group. The inverse to $R$ will
be associated with the reverse elementary braid on two strands.
Next, we define a representation $\tau$ of the Artin braid group to
the automorphism of $\mathcal{V}^{\otimes
m}$ by
\begin{equation}\label{ABG}
    \tau(b_{i})=\mathcal{I}\otimes\cdots\otimes\mathcal{I}\otimes R\otimes
    \mathcal{I}\otimes\cdots\otimes\mathcal{I},
\end{equation}
where $R$ act on  $\mathcal{V}_{i}\otimes\mathcal{V}_{i+1}$. This
equation describes a representation of the braid group if $R$
satisfies the Yang-Baxter equation and is also invertible. Moreover,
this representation of braid group is unitary if $R$ is also unitary
operator. Thus $R$ being unitary indicated that this operator can
performs topological entanglement and it also can be considers as
quantum gate. It has been show in \cite{kauf} that $R$ can also
perform quantum entanglement by acting on qubits states. An
associative unital algebra $A$  with homomorphism
$\Delta:A\longrightarrow A\otimes A$ is called co-multiplication and
also co-associative, that is for all $a\in A$ we have
\begin{equation}
(\mathcal{I}\otimes \Delta)(\Delta(a))=(\Delta\otimes \mathcal{I})(\Delta(a)).
\end{equation}
We can also construct another co-product $\Delta^{'}=\sigma
\circ\Delta$, where, $\sigma$ is a permutation defined by
$\sigma(a_{1}\otimes a_{2})=\sigma(a_{2}\otimes a_{1})$. Next, let
$\mathcal{R}\in A$ such that
$(\Delta^{'}(a))=\mathcal{R}(\Delta(a))\mathcal{R}^{-1}$. The Hopf
algebra $A$ is called quasi-triangular if
\begin{equation}
(\Delta\otimes \mathcal{I})(\mathcal{R})=\mathcal{R}_{13}\mathcal{R}_{23},
~\text{and}~(\mathcal{I}\otimes \Delta)(\mathcal{R})=\mathcal{R}_{13}\mathcal{R}_{12},
\end{equation}
where  $\mathcal{R}_{12}=\sum_{i}a_{i}\otimes b_{i}\otimes \mathcal{I}$,
$\mathcal{R}_{12}=\sum_{i}\mathcal{I}\otimes a_{i}\otimes b_{i}$, and
 $\mathcal{R}_{13}=\sum_{i}a_{i}\otimes \mathcal{I}\otimes b_{i}$.
  Moreover, $R$ is called a universal $\mathcal{R}$-matrix and satisfies the following relation
\begin{equation}
\mathcal{R}_{12}\mathcal{R}_{13}R_{23}=\mathcal{R}_{23}\mathcal{R}_{13}\mathcal{R}_{12}.
\end{equation}
This equation implies that the matrix $R=\Pi \mathcal{R}$, where $\Pi$ is a permutation operator satisfies quantum Yang-Baxter equation \ref{YB}.
\section{Concurrence classes for general  multipartite  states}\label{con}
In this section we will review the construction of concurrence classes based on orthogonal
 complement of a POVM on quantum phase.
Let us denote a general, multipartite quantum system with $m$
subsystems by
$\mathcal{Q}=\mathcal{Q}_{m}(N_{1},N_{2},\ldots,N_{m})$ $
=\mathcal{Q}_{1}\mathcal{Q}_{2}\cdots\mathcal{Q}_{m}$, consisting
of a state
\begin{equation}\label{Mstate}
\ket{\Psi}=\sum^{N_{1}}_{j_{1}=1}\sum^{N_{2}}_{j_{2}=1}\cdots\sum^{N_{m}}_{j_{m}=1}
\alpha_{j_{1}j_{2}\cdots j_{m}} \ket{j_{1}j_{2}\cdots j_{m}}
\end{equation}
 and, let
$\rho_{\mathcal{Q}}=\sum^{\mathrm{N}}_{n=1}p_{n}\ket{\Psi_{n}}\bra{\Psi_{n}}$,
for all $0\leq p_{n}\leq 1$ and $\sum^{\mathrm{N}}_{n=1}p_{n}=1$,
denote a density operator acting on the Hilbert space $
\mathcal{H}_{\mathcal{Q}}=\mathcal{H}_{\mathcal{Q}_{1}}\otimes
\mathcal{H}_{\mathcal{Q}_{2}}\otimes\cdots\otimes\mathcal{H}_{\mathcal{Q}_{m}},
$ where the dimension of the $j$th Hilbert space is given  by
$N_{j}=\dim(\mathcal{H}_{\mathcal{Q}_{j}})$. Moreover, let us
introduce a complex conjugation operator $\mathcal{C}_{m}$ that
acts on a general state $\ket{\Psi}$ of a multipartite state as
$
\mathcal{C}_{m}\ket{\Psi}=\sum^{N_{1}}_{j_{1}=1}\sum^{N_{2}}_{j_{2}=1}\cdots\sum^{N_{m}}_{j_{m}=1}
\alpha^{*}_{j_{1}j_{2}\cdots j_{m}}
\ket{j_{1}j_{2}\cdots j_{m}}
$.
 The
density operator $\rho_{\mathcal{Q}}$ is said to be fully
separable, which we will denote by $\rho^{sep}_{\mathcal{Q}}$,
with respect to the Hilbert space decomposition, if it can  be
written as $ \rho^{sep}_{\mathcal{Q}}=\sum^\mathrm{N}_{n=1}p_{n}
\bigotimes^m_{j=1}\rho^{n}_{\mathcal{Q}_{j}}$,
$\sum^\mathrm{N}_{n=1}p_{n}=1 $,
 for some positive integer $\mathrm{N}$, where $p_{n}$ are positive real
numbers and $\rho^{n}_{\mathcal{Q}_{j}}$ denote a density operator
on Hilbert space $\mathcal{H}_{\mathcal{Q}_{j}}$. If
$\rho^{p}_{\mathcal{Q}}$ represents a pure state, then the quantum
system is fully separable if $\rho^{p}_{\mathcal{Q}}$ can be
written as
$\rho^{sep}_{\mathcal{Q}}=\bigotimes^m_{j=1}\rho_{\mathcal{Q}_{j}}$,
where $\rho_{\mathcal{Q}_{j}}$ is a density operator on
$\mathcal{H}_{\mathcal{Q}_{j}}$. If a state is not separable, then
it is called an entangled state.

Now,
a general and symmetric POVM in a single $N_{j}$-dimensional
Hilbert space $\mathcal{H}_{\mathcal{Q}_{j}}$ is given by
\begin{equation}
\Delta(\varphi_{1_j,2_j},\ldots,\varphi_{1_j,N_j},
\varphi_{2_j,3_j},\ldots,\varphi_{N_{j}-1,N_{j}})=
\sum^{N_{j}}_{l_{j}}\sum^{N_{j}}_{k_{j}=1}
e^{i\varphi_{k_{j},l_{j}}}\ket{k_{j}}\bra{l_{j}},
\end{equation}
where $\ket{k_{j}}$ and $\ket{l_{j}}$ are the basis vectors in
$\mathcal{H}_{\mathcal{Q}_j}$ and the quantum phases satisfy the
following relation $ \varphi_{k_{j},l_{j}}=
-\varphi_{l_{j},k_{j}}(1-\delta_{k_{j} l_{j}})$, see Ref.
\cite{Hosh3}. The POVM is a function of the $N_{j}(N_{j}-1)/2$
phases
$(\varphi_{1_j,2_j},\ldots,\varphi_{1_j,N_j},\varphi_{2_j,3_j},\ldots,\varphi_{N_{j}-1,N_{j}})$.
It is now possible to form a POVM of a multipartite system by
simply forming the tensor product
\begin{eqnarray}\label{POVM}\nonumber
\Delta_\mathcal{Q}(\varphi_{k_{1},l_{1}},\ldots,
\varphi_{k_{m},l_{m}})&=&
\Delta_{\mathcal{Q}_{1}}(\varphi_{k_{1},l_{1}})\otimes\cdots
\otimes
\Delta_{\mathcal{Q}_{m}}(\varphi_{k_{m},l_{m}}),
\end{eqnarray}
where, e.g., $\varphi_{k_{1},l_{1}}$ is the set of
POVMs phase associated with subsystems $\mathcal{Q}_{1}$, for all
$k_{1},l_{1}=1,2,\ldots,N_{1}$, where we need only to consider
when $l_{1}>k_{1}$.
Let us now construct concurrence classes for general
 multipartite states
$\mathcal{Q}_{m}(N_{1},\ldots,N_{m})$. In order to simplify our
presentation,  we will use $\Omega_{m}=\varphi_{k_{1},l_{1}},\ldots,
\varphi_{k_{m},l_{m}}$ as an abstract multi-index notation. The
unique structure of our POVM enables us to distinguish different
classes of multipartite states, which are inequivalent under LOCC
operations. In the $m$-partite case, the  off-diagonal elements of
the matrix corresponding to
\begin{equation}
\widetilde{\Delta}_\mathcal{Q}(\Omega_{m})=
\widetilde{\Delta}_{\mathcal{Q}_{1}}(\varphi_{k_{1},l_{1}})
\otimes\cdots
\otimes\widetilde{\Delta}_{\mathcal{Q}_{m}}(\varphi_{k_{m},l_{m}}),
\end{equation}
 have phases that are sum
or differences of phases originating from two and $m$ subsystems.
That is, in the later case the phases of
$\widetilde{\Delta}_\mathcal{Q}(\varphi_{k_{1},l_{1}},\ldots,
\varphi_{k_{m},l_{m}})$ take the form
$(\varphi_{k_{1},l_{1}}\pm\varphi_{k_{2},l_{2}}
\pm\ldots\pm\varphi_{k_{m},l_{m}})$ and
identification of these joint phases makes our classification
possible. Thus, we  can define linear operators for  the
$EPR_{\mathcal{Q}_{r_{1}}\mathcal{Q}_{r_{2}}}$ class based on our
POVM which are sum and difference of phases of two subsystems,
i.e., $(\varphi_{k_{r_{1}},l_{r_{1}}}
\pm\varphi_{k_{r_{2}},l_{r_{2}}})$. That is,
for the $EPR_{\mathcal{Q}_{r_{1}}\mathcal{Q}_{r_{2}}}$  class we
have
\begin{eqnarray}\nonumber
 \widetilde{\Delta}^{EPR}_{\mathcal{Q}_{r_{1},r_{2}}}(\Omega_{m})
&=&\mathcal{I}_{N_{1}} \otimes\cdots
\otimes\widetilde{\Delta}_{\mathcal{Q}_{r_{1}}}
(\varphi^{\frac{\pi}{2}}_{k_{r_{1}},l_{r_{1}}})
\otimes\cdots\otimes \widetilde{\Delta}_{\mathcal{Q}_{r_{2}}}
(\varphi^{\frac{\pi}{2}}_{k_{r_{2}},l_{r_{2}}})\\&&\otimes\cdots\otimes
\mathcal{I}_{N_{m}}.
\end{eqnarray}
 For the $GHZ^{m}$ class, we define the linear
operators based on our POVM which are sum and difference of phases
of $m$-subsystems, i.e.,
$(\varphi_{k_{r_{1}},l_{r_{1}}}
\pm\varphi_{k_{r_{2}},l_{r_{2}}}\pm
\ldots\pm\varphi_{k_{m},l_{m}})$. That is, for the
$GHZ^{m}$ class we have
\begin{eqnarray}
\widetilde{\Delta}^{
GHZ^{m}}_{\mathcal{Q}_{r_{1},r_{2}}}(\Omega_{m})
&=&\widetilde{\Delta}_{\mathcal{Q}_{r_{1}}}
(\varphi^{\frac{\pi}{2}}_{k_{r_{1}},l_{r_{1}}})
\otimes\widetilde{\Delta}_{\mathcal{Q}_{r_{2}}}
(\varphi^{\frac{\pi}{2}}_{k_{r_{2}},l_{r_{2}}})
\otimes\\\nonumber&&\widetilde{\Delta}_{\mathcal{Q}_{r_{3}}}
(\varphi^{\pi}_{k_{r_{3}},l_{r_{3}}})
\otimes\cdots
\otimes\widetilde{\Delta}_{\mathcal{Q}_{m}}
(\varphi^{\pi}_{k_{r_{m}},l_{r_{m}}}),
\end{eqnarray}
where by choosing
$\varphi^{\pi}_{k_{j},l_{j}}=\pi$ for all
$k_{j}<l_{j}, ~j=1,2,\ldots,m$, we get an operator which has the
structure of Pauli operator $\sigma_{x}$ embedded in a
higher-dimensional Hilbert space and coincides with $\sigma_{x}$
for a single-qubit. There are $C(m,2)$ linear operators for the
$GHZ^{m}$ class and the set of these operators gives the $GHZ^{m}$
class concurrence.


\section{Multipartite quantum gate entangler}\label{gen}
In this section we will construct quantum gates entanglers for W and
GHZ classes of multipartite quantum states. Our construction is
based on definition of W and GHZ concurrence classes for general
multipartite states \cite{Hosh5,Hosh6}.
Let us consider quantum system $\mathcal{Q}^{p}_{m}(N,N,\ldots,N)$,
where $N_{1}=N_{2}=\cdots=N_{m}=N=2$ and $p$ indicates that we
consider a quantum system in  pure state. Then, for a multipartite
state a topological unitary transformation $\mathcal{R}_{N^{m}\times
N^{m}}$ that create multipartite entangled state is defined by
$\mathcal{R}_{N^{m}\times N^{m}}=\mathcal{R}^{d}_{N^{m}\times
N^{m}}+\mathcal{R}^{ad}_{N^{m}\times N^{m}}$, where
\begin{equation}
\mathcal{R}^{a}_{N^{m}\times N^{m}}=\text{diag}(\alpha_{11\cdots1},0,\ldots,0,\alpha_{NN\cdots N})
\end{equation}
 is a
diagonal matrix and
\begin{equation}\mathcal{R}^{ad}_{N^{m}\times N^{m}}=\text{antidiag}(0,\alpha_{NN\cdots N-1},\alpha_{NN\cdots N-1N},
\ldots,\alpha_{1\cdots2 1},\alpha_{1\cdots1 2},0)
\end{equation}
 is an anti-diagonal matrix.
Now, we have the following proposition for multipartite states.
\begin{prop} \label{epr}If elements of $\mathcal{R}_{N^{m}\times N^{m}}$
satisfy
\begin{eqnarray}\nonumber\label{conEPR}
   &&
    \langle \Psi\ket{\widetilde{\Delta}^{
EPR}_{\mathcal{Q}_{r_{1},r_{2}}}(\Omega_{m})
\mathcal{C}_{m}\Psi}\neq0
\end{eqnarray}
then the state
$\mathcal{R}_{N^{m}\times N^{m}}(\ket{\psi}\otimes\ket{\psi}\otimes\cdots\otimes\ket{\psi})$, with
$\ket{\psi}=\ket{1}+\ket{2}+\ldots+\ket{N}$ is a EPR or W class entangled.
\end{prop}
The proof  follows from the  construction of
$\mathcal{R}_{N^{m}\times N^{m}}$ which is based on separable
elements of multipartite states given by EPR class of operator
presented in \cite{Hosh5,Hosh6}.
\begin{prop}\label{ghz} If elements of $\mathcal{R}_{N^{m}\times N^{m}}$
satisfy
\begin{eqnarray}
  &&\langle \Psi\ket{\widetilde{\Delta}^{
GHZ^{m}}_{\mathcal{Q}_{r_{1},r_{2}}
}(\Omega_{m})\mathcal{C}_{m}\Psi}\neq0
\end{eqnarray}
then the state
$\mathcal{R}_{N^{m}\times N^{m}}(\ket{\psi}\otimes\ket{\psi}\otimes\cdots\otimes\ket{\psi})$, with
$\ket{\psi}=\ket{1}+\ket{2}+\ldots+\ket{N}$ is a GHZ class entangled.
\end{prop}
The proof  follows from the  construction of
$\mathcal{R}_{N^{m}\times N^{m}}$ which is based on  separable
elements of multipartite states defined by  GHZ class of operator
\cite{Hosh5,Hosh6}.
 Note that this operator is a quantum gate
entangler since
\begin{eqnarray}
\tau_{N^{m}\times N^{m}}&=&\mathcal{R}_{N^{m}\times
N^{m}}\mathcal{P}_{N^{m}\times N^{m}}\\\nonumber&=&
\mathrm{diag}(\alpha_{1\cdots1 1},\alpha_{1\cdots1
2},\ldots,\alpha_{N\cdots N N})
\end{eqnarray}
 is a $N^{m}\times N^{m}$
phase gate and $\mathcal{P}_{N^{m}\times N^{m}}$ is $N^{m}\times
N^{m}$ a "swap gate". However, we need to imposed some constraints
on the parameters $\alpha_{k_{1}k_{2}\cdots k_{m}}$ to ensure that
our $\mathcal{R}$ is unitary. This is our main result and in
following section
 we will illustrate it by some examples.

\section{Quantum gate entangler for three-partite states}\label{exa}
In this section, we will construct a quantum gate entangler for
three-partite  quantum systems. Let us consider quantum system
$\mathcal{Q}^{p}_{m}(N,N,N)$, where $N_{1}=N_{2}=N_{3}=N=2$. Then,
for a general three-partite state a topological unitary
transformation $\mathcal{R}_{N^{3}\times N^{3}}$ that create
multipartite entangled state is defined by $\mathcal{R}_{N^{3}\times
N^{3}}=\mathcal{R}^{d}_{N^{3}\times
N^{3}}+\mathcal{R}^{ad}_{N^{3}\times N^{3}}$, where
\begin{equation}
\mathcal{R}^{a}_{N^{3}\times
N^{3}}=\text{diag}(\alpha_{111},0,\ldots,0,\alpha_{222})
\end{equation}
 is a
diagonal matrix and
\begin{equation}\mathcal{R}^{ad}_{N^{3}\times N^{3}}=\text{antidiag}(0,\alpha_{221},\alpha_{212},
\ldots,\alpha_{121},\alpha_{11 2},0)
\end{equation}
 is an anti-diagonal matrix.
Moreover, for three-partite states we have the following EPR class operators
\begin{eqnarray}\label{W3}\nonumber
 \widetilde{\Delta}^{
EPR}_{\mathcal{Q}_{1,2}}(\Omega_{3})
&=&\widetilde{\Delta}_{\mathcal{Q}_{1}}
(\varphi^{\frac{\pi}{2}}_{k_{1},l_{1}})
\otimes\widetilde{\Delta}_{\mathcal{Q}_{2}}
(\varphi^{\frac{\pi}{2}}_{k_{2},l_{2}})\otimes
\mathcal{I}_{N_{3}}.
\end{eqnarray}
$
 \widetilde{\Delta}^{
EPR}_{\mathcal{Q}_{1,3}}(\Omega_{3}) $ and $
 \widetilde{\Delta}^{
EPR}_{\mathcal{Q}_{2,3}} (\Omega_{3})$ are defined  in a similar
way. Now, based on proposition  \ref{epr}, If elements of
$\mathcal{R}_{N^{3}\times N^{3}}$ satisfies
\begin{equation}\label{conW3}
 \langle \Psi\ket{\widetilde{\Delta}^{
EPR}_{\mathcal{Q}_{1,2}}(\Omega_{3})
\mathcal{C}_{3}\Psi}=\sum^{N_{1}}_{l_{1}>k_{1}=1}
\sum^{N_{2}}_{l_{2}>k_{2}=1} \{\sum^{N_{3}}_{k_{3}=l_{3}=1}(
\alpha_{k_{1}l_{2}k_{3}}\alpha_{l_{1}k_{2}l_{3}}-\alpha_{k_{1}k_{2}k_{3}}\alpha_{l_{1}l_{2}l_{3}}
) \}\neq0,
\end{equation}
\begin{equation}\label{conW3}
\langle \Psi\ket{\widetilde{\Delta}^{
EPR}_{\mathcal{Q}_{1,3}}(\Omega_{3})
\mathcal{C}_{3}\Psi}=\sum^{N_{1}}_{l_{1}>k_{1}=1}
\sum^{N_{3}}_{l_{3}>k_{3}=1}\{
\sum^{N_{2}}_{k_{2}=l_{2}=1}(\alpha_{k_{1}k_{2}l_{3}}\alpha_{l_{1}l_{2}k_{3}}
-\alpha_{k_{1}k_{2}k_{3}}\alpha_{l_{1}l_{2}l_{3}}
) \}\neq0,
\end{equation}
\begin{equation}\label{conW3}
    \langle \Psi\ket{\widetilde{\Delta}^{
EPR}_{\mathcal{Q}_{2,3}}(\Omega_{3})
\mathcal{C}_{3}\Psi}=\sum^{N_{2}}_{l_{2}>k_{2}=1}
\sum^{N_{3}}_{l_{3}>k_{3}=1} \{\sum^{N_{1}}_{k_{1}=l_{1}=1}(
\alpha_{k_{1}k_{2}l_{3}}\alpha_{l_{1}l_{2}k_{3}}-\alpha_{k_{1}k_{2}k_{3}}\alpha_{l_{1}l_{2}l_{3}}
)\}\neq0,
\end{equation}
 then the state
$\mathcal{R}_{N^{3}\times
N^{3}}(\ket{\psi}\otimes\ket{\psi}\otimes\ket{\psi})$, with
$\ket{\psi}=\ket{1}+\ket{2}$ is a  W class entangled.

 The second class of three-partite
state that we would like to consider is the $GHZ^{3}$ class. For
this class, we have
\begin{eqnarray}\nonumber
\widetilde{\Delta}^{ GHZ^{3}}_{\mathcal{Q}_{1,2}
}(\Omega_{3})&=& \widetilde{\Delta}_{\mathcal{Q}_{1}}
(\varphi^{\frac{\pi}{2}}_{k_{1},l_{1}})
\otimes\widetilde{\Delta}_{\mathcal{Q}_{2}}
(\varphi^{\frac{\pi}{2}}_{k_{2},l_{2}})\otimes
\widetilde{\Delta}_{\mathcal{Q}_{3}}
(\varphi^{\pi}_{k_{3},l_{3}}).
\end{eqnarray}
$
 \widetilde{\Delta}^{
GHZ^{3}}_{\mathcal{Q}_{1,3}(\Omega_{3})} $ and $
 \widetilde{\Delta}^{
GHZ^{3}}_{\mathcal{Q}_{2,3}}(\Omega_{3}) $ are defined in a similar
way. Now, based on proposition  \ref{ghz}, If elements of
$\mathcal{R}_{N^{3}\times N^{3}}$ satisfies
\begin{eqnarray}\label{conGHZ3}
  &&\langle \Psi\ket{\widetilde{\Delta}^{
GHZ^{3}}_{\mathcal{Q}_{1,2}}(\Omega_{3})
\mathcal{C}_{3}\Psi}=\sum^{N_{1}}_{l_{1}>k_{1}=1}
\sum^{N_{2}}_{l_{2}>k_{2}=1}
\sum^{N_{3}}_{l_{3}>k_{3}=1}
\\\nonumber&& \{
\alpha_{k_{1}l_{2}l_{3}}\alpha_{l_{1}k_{2}k_{3}}+
\alpha_{k_{1}l_{2}k_{3}}\alpha_{l_{1}k_{2}l_{3}}
-\alpha_{k_{1}k_{2},l_{3}}\alpha_{l_{1}l_{2}k_{3}}
-\alpha_{k_{1}k_{2},k_{3}}\alpha_{l_{1}l_{2}l_{3}}
\}\neq0,
\end{eqnarray}
\begin{eqnarray}\label{conGHZ3}
  &&\langle \Psi\ket{\widetilde{\Delta}^{
GHZ^{3}}_{\mathcal{Q}_{1,3}}(\Omega_{3})
\mathcal{C}_{3}\Psi}=\sum^{N_{1}}_{l_{1}>k_{1}=1}
\sum^{N_{2}}_{l_{2}>k_{2}=1}
\sum^{N_{3}}_{l_{3}>k_{3}=1}
\\\nonumber&& \{
\alpha_{k_{1}l_{2}l_{3}}\alpha_{l_{1}k_{2}k_{3}}-
\alpha_{k_{1}l_{2}k_{3}}\alpha_{l_{1}k_{2}l_{3}}
+\alpha_{k_{1}k_{2}l_{3}}\alpha_{l_{1}l_{2}k_{3}}
-\alpha_{k_{1}k_{2}k_{3}}\alpha_{l_{1}l_{2}l_{3}}
\}\neq0,
\end{eqnarray}
\begin{eqnarray}\label{conGHZ3}
  &&\langle \Psi\ket{\widetilde{\Delta}^{
GHZ^{3}}_{\mathcal{Q}_{2,3}}(\Omega_{3})
\mathcal{C}_{3}\Psi}=\sum^{N_{1}}_{l_{1}>k_{1}=1}
\sum^{N_{2}}_{l_{2}>k_{2}=1}
\sum^{N_{3}}_{l_{3}>k_{3}=1}
\\\nonumber&& \{
-\alpha_{k_{1}l_{2}l_{3}}\alpha_{l_{1}k_{2}k_{3}}+
\alpha_{k_{1}l_{2}k_{3}}\alpha_{l_{1}k_{2}l_{3}}
+\alpha_{k_{1}k_{2}l_{3}}\alpha_{l_{1}l_{2}k_{3}}
-\alpha_{k_{1}k_{2}k_{3}}\alpha_{l_{1}l_{2}l_{3}}
\}\neq0,
\end{eqnarray}
 then the state
$\mathcal{R}_{N^{3}\times
N^{3}}(\ket{\psi}\otimes\ket{\psi}\otimes\ket{\psi})$, with
$\ket{\psi}=\ket{1}+\ket{2}$ is a GHZ class entangled.
Thus, we have constructed two different classes of quantum gates
entanglers for multipartite
 states based on construction of the concurrence. This result is the first step toward the
  construction of quantum gates entangler  for specific classes of multipartite  entangled states.
  These set of operators are very important for design of entangled based quantum computer and needs
  further investigation.

\begin{flushleft}
\textbf{Acknowledgments:} The  author gratefully acknowledges the
financial support of the Japan Society for the Promotion of Science
(JSPS).
\end{flushleft}


\end{document}